%% file: be_pp_v4.tex
\documentclass[twocolumn,aps,pra,floatfix,superscriptaddress,notitlepage]{revtex4-1}
\usepackage[utf8]{inputenc}
\usepackage[OT1]{fontenc}
\usepackage{amsmath}
\usepackage{amsfonts}
\usepackage{amssymb}
\usepackage{bm}
\usepackage{graphicx}
\usepackage[left=2cm,right=2cm,top=2cm,bottom=2cm]{geometry}
\usepackage{longtable,booktabs}
\allowdisplaybreaks
\usepackage{dcolumn}
\usepackage{siunitx}
\usepackage{multirow}
\usepackage{bigdelim}

\usepackage{color}
\definecolor{BLUE}{rgb}{0.0,0.0,1.0}

\usepackage{soul}
\soulregister\cite7
\soulregister\ref7

\newcommand{\be}{\begin{eqnarray}}
\newcommand{\ee}{\end{eqnarray}}

\begin{document}

\title{QED calculations of intra-$L$-shell doubly excited states in Be-like ions}

\author{A.~V.~Malyshev}
\affiliation{Department of Physics, St.~Petersburg State University, Universitetskaya 7-9, 199034 St.~Petersburg, Russia  
\looseness=-1}
\affiliation{Petersburg Nuclear Physics Institute named by B.P. Konstantinov of National Research Center ``Kurchatov Institute'', Orlova roscha 1, 188300 Gatchina, Leningrad region, Russia}

\author{Y.~S.~Kozhedub}
\affiliation{Department of Physics, St.~Petersburg State University, Universitetskaya 7-9, 199034 St.~Petersburg, Russia  
\looseness=-1}

\author{V.~M.~Shabaev}
\affiliation{Department of Physics, St.~Petersburg State University, Universitetskaya 7-9, 199034 St.~Petersburg, Russia  
\looseness=-1}
\affiliation{Petersburg Nuclear Physics Institute named by B.P. Konstantinov of National Research Center ``Kurchatov Institute'', Orlova roscha 1, 188300 Gatchina, Leningrad region, Russia}

\author{I.~I.~Tupitsyn}
\affiliation{Department of Physics, St.~Petersburg State University, Universitetskaya 7-9, 199034 St.~Petersburg, Russia  
\looseness=-1}


\begin{abstract}

The rigorous QED approach is employed to calculate the energies of the $2p2p\,^3P_{0,1,2}$, $2p2p\,^1D_2$, and $2p2p\,^1S_0$ states of selected Be-like highly charged ions over a wide range of nuclear-charge numbers, $18 \leqslant Z \leqslant 92$. Combined with the previously reported energies of the $2s2p \, ^3P_{0,1,2}$ and $2s2p \, ^1P_1$ states [A.~V.~Malyshev \textit{et al.}, Phys. Rev. A {\bf 110}, 062824 (2024)], the obtained results are used to study various intra-$L$-shell transition energies. Strong level mixing, caused by the proximity of states with the same symmetry, is overcome by means of the QED perturbation theory for quasi-degenerate levels. The applied approach merges a rigorous perturbative QED treatment up to the second order with the consideration of electron-electron correlation contributions of the third and higher orders evaluated within the Breit approximation. The higher-order screened QED effects are estimated using the model-QED-operator approach. The nuclear-recoil and nuclear-polarization effects are also taken into account. The obtained predictions represent the most accurate theoretical description of the electronic structure of Be-like ions to date and demonstrate good agreement with available experimental data.
 
\end{abstract}


\maketitle


\section{Introduction \label{sec:0}}

Highly charged ions (HCIs) provide an ideal platform for conducting a variety of fundamental studies~\cite{Andreev:2005:243002, Shabaev:2006:253002, CrespoLopez-Urrutia:2008:111, Berengut:2010:120801, Sturm:2014:467, Yerokhin:2016:100801, Oreshkina:2017:030501_R, Kozlov:2018:045005, Yerokhin:2020:012502, Debierre:2020:135527, Blaum:2020:014002, Rehbehn:2021:L040801, Shabaev:2022:043001, King:2022:43, Debierre:2022:062801}. This is justified by two key factors. First, all non-trivial relativistic and quantum-electrodynamic (QED) effects in HCIs are significantly enhanced compared to those observed in light atoms. Second, these effects are not obscured by the uncertainty of electron-electron correlation calculations, as is typically the case in heavy neutral systems. 

Among the fundamental investigations involving HCIs, tests of bound-state QED hold a central position~\cite{Sapirstein:2008:25, Cheng:2008:33, Beiersdorfer:2010:074032, Volotka:2012:073001, Shabaev:2018:60, Indelicato:2019:232001}. While it is undisputed that QED is a well-established theory of light and matter interaction, the aforementioned tests actually serve to validate based-on-QED theoretical approaches developed for the evaluation of various atomic properties. It is essential that $\alpha Z$-expansion methods ($\alpha$ being the fine-structure constant and $Z$ being the nuclear-charge number) designed for light systems~\cite{Caswell:1986:437, Pachucki:1998:5123, Korobov:2025:43} are not applicable to HCIs. Instead, novel \textit{ab initio} techniques, which are nonperturbative in the nuclear-strength parameter $\alpha Z$~\cite{Mohr:1998:227, TTGF}, must be developed for accurate theoretical treatment in this regime.

When discussing the electronic structure, textbook examples of the bound-state-QED tests include comparisons of theoretical predictions and high-precision measurements of the ground-state Lamb shift in H-like uranium~\cite{Stoehlker:2000:3109, Gumberidze:2005:223001} and the $2p \, ^2P_{1/2} \rightarrow 2s \, ^2S_{1/2}$ transition energy in Li-like uranium~\cite{Schweppe:1991:1434, Brandau:2003:073202, Beiersdorfer:2005:233003}. For the sake of brevity, we omit the closed $1s^2$ shell from the state designations here and in what follows, if this does not lead to misunderstandings. The QED description of these ions obviously shares some common aspects. All one-electron contributions can be treated using the same approaches. For instance, a complete evaluation of all one-electron two-loop contributions to all orders in $\alpha Z$, which is anticipated in the near future (see Refs.~\cite{Yerokhin:2025:042820, Volkov:2025:preprint_03284} for recent progress), will evidently affect the accuracy of theoretical predictions not only for these systems, but also for other charge states of heavy few-electron ions. Meanwhile, compared to H-like systems, the consideration of Li-like ions presents a qualitatively new challenge for theory, as it requires the rigorous treatment of electron correlation effects within a QED framework. However, the $2s \, ^2S_{1/2}$ and $2p \, ^2P_{1/2,3/2}$ states of Li-like ions provide an example of well-separated energy levels. All these states have different symmetries and, therefore, are not mixed by interelectronic interaction. For this reason, the conventional QED perturbation theory (PT) formulated for single levels can be used for their accurate description~\cite{Kozhedub:2010:042513, Sapirstein:2011:012504, Yerokhin:2025:preprint}. The level mixing becomes relevant when considering, e.g., the singly excited $1s2p \, ^3P_1$ and $1s2p \, ^1P_1$ states in He-like ions. In such cases, a more sophisticated QED PT for quasi-degenerate levels has to be applied~\cite{Artemyev:2005:062104, Malyshev:2019:010501_R, Kozhedub:2019:062506, Yerokhin:2022:022815}. Some time ago, there was a discussion in the literature about a possible discrepancy between theoretical predictions and measurements of x-ray transition lines in He-like ions~\cite{Chantler:2012:153001, Chantler:2014:123037}, which was followed, however, by a series of experimental works supporting the QED theory of HCIs, see Refs.~\cite{Kubicek:2014:032508, Epp:2015:020502_R, Beiersdorfer:2015:032514, Machado:2018:032517, Loetzsch:2024:673} and references therein. Nevertheless, further comprehensive testing of the underlying QED approaches is needed.

In this respect, Be-like ions represent another important example of few-electron systems where the mixing of levels is crucial. Previously, numerous attempts were undertaken to treat Be-like ions within different methods~\cite{Armstrong:1976:1114, Cheng:1979:111, Edlen:1983:51, Edlen:1985:86, Lindroth:1992:2771, Marques:1993:929, Zhu:1994:3818, Safronova:1996:4036, Chen:1997:166, Santos:1998:149, Cheng:2000:054501, Safronova:2000:1213, Majumder:2000:042508, Dong:2001:294, Gu:2005:267, Ho:2006:022510, Cheng:2008:052504, Yerokhin:2014:022509, Yerokhin:2015:054003, Wang:2015:16, Kaygorodov:2019:032505}. However, as far as we know, all these studies have incorporated the QED effect, at best, within some one-electron (first-order) or semiempirical approximations. This has resulted in a substantial scatter of the obtained theoretical predictions. In Refs.~\cite{Malyshev:2014:062517, Malyshev:2015:012514}, the ground state of Be-like ions was treated within the QED PT for single levels, taking into account all QED effects up to the second order. However, the uncertainty associated with the level mixing was underestimated. Later, in Ref.~\cite{Malyshev:2021:183001}, we showed that the strong interplay between the QED and correlation effects in Be-like HCIs can be properly addressed only within the framework of the QED perturbative approach for quasidegenerate levels. Pilot calculations based on the developed method~\cite{Malyshev:2021:183001, Malyshev:2021:652, Malyshev:2023:042806} were extended in Ref.~\cite{Malyshev:2024:062824} to cover a wide range of Be-like HCIs, from argon to uranium. These studies provided a detailed analysis of the energies of singly excited $2s2p \, ^3P_{0,1,2}$ and $2s2p \, ^1P_1$ states as well as the $2s2p \, ^3P_1 \rightarrow 2s2p \, ^3P_0$ and $2s2p \, ^3P_2 \rightarrow 2s2p \, ^3P_1$ transition energies. The theoretical predictions show good agreement with most available experimental data~\cite{Widing:1975:L33, Dere:1978:1062, Denne:1989:1488, Denne:1989:3702, Beiersdorfer:1993:3939, Beiersdorfer:1995:2693, Feili:2005:48, Beiersdorfer:2005:233003, Bernhardt:2015:144008, Loetzsch:2024:673}, thereby strongly confirming the reliability of the developed approach. Along with this, noticeable discrepancies with some measurements \cite{Trabert:2003:042501, Beiersdorfer:1998:1944} were identified, which highlights the need for new and more accurate experimental studies involving Be-like HCIs.
 
The present work is a natural extension of Ref.~\cite{Malyshev:2024:062824}. Namely, we investigate the doubly excited $2p2p\,^3P_{0,1,2}$, $2p2p\,^1D_2$, and $2p2p\,^1S_0$ states of Be-like HCIs and evaluate their energies relative to the $2s2s \, ^1S_0$ ground state. A number of relevant transition energies is studied as well. As in Ref.~\cite{Malyshev:2024:062824}, an accurate analysis of uncertainties associated with uncalculated effects is performed, and the obtained results are compared with the previous theoretical predictions and available measurements. The goal of this study is to complete
\textit{ab initio} calculations of all intra-$L$-shell excitations in Be-like HCIs and establish a benchmark for future high-precision experimental and theoretical investigations.


\section{Theoretical approach and computational details \label{sec:1}}

The \textit{ab initio} method employed in the present work to treat the intra-$L$-shell excitations in Be-like HCIs was generally formulated in our earlier works~\cite{Kozhedub:2019:062506, Malyshev:2021:183001, Malyshev:2021:652, Malyshev:2023:042806, Malyshev:2024:062824}. Here, we only provide a brief overview of its main features and refer interested readers to these papers for more details. In addition, a minor modification in the method that concerns the one-electron two-loop contributions is also discussed below.

Our approach is based on the QED PT formulated within the Furry picture~\cite{Furry:1951:115} in the framework of the two-time Green’s function (TTGF) method~\cite{TTGF}. The zeroth-order approximation is determined by the one-electron Dirac equation, which along with the Coulomb potential of an extended nucleus, $V_{\rm nucl}$, includes also a local spherically symmetric screening potential, $V_{\rm scr}$. The latter one is added to partly take into account the interelectronic-interaction effects from the very beginning. When incorporating the screening potential into the zeroth-order Hamiltonian, the perturbation series is rearranged. We note that in the corresponding formulation of PT, the counterterm $\delta V=-V_{\rm scr}$ must be treated perturbatively. In the present calculations, we adopt the local Dirac-Fock (LDF) screening potential~\cite{Shabaev:2005:062105} as the main one. All the results discussed below are obtained using this potential. However, in order to keep under control the accuracy and to analyze the convergence of PT, we also perform calculations starting from the core-Hartree (CH) potential induced by the closed $1s^2$ shell, as well as ones without including any screening potential at all.

The applied QED PT involves all contributions up to the second order, which corresponds to the current state of the art in this field. All first-order and many-electron second-order terms are rigorously calculated. The situation with the one-electron second-order (two-loop) contributions, whose evaluation to all orders in $\alpha Z$ represents a very complicated task, is currently as follows. At present, a part of these contributions has still been considered only within the free-loop approximation~\cite{Yerokhin:2008:062510}. However, significant progress has been achieved over the past year. First, the convergence-acceleration approach by Sapirstein and Cheng~\cite{Sapirstein:2023:042804}, originally proposed for the first-order self-energy contribution, has been successfully extended to treat two-loop self-energy diagrams~\cite{Yerokhin:2024:251803, Yerokhin:2025:042820}. Second, two-loop vacuum-polarization contributions, which were one of the main sources of theoretical uncertainty, have been evaluated~\cite{Volkov:2025:preprint_03284, Volkov:2025:preprint_02957}. In our previous works, for the one-electron two-loop contributions, we used the results summarized in Refs.~\cite{Yerokhin:2015:033103, Yerokhin:2018:052509}. In the present study, we incorporate the most recent updates reported in Refs.~\cite{Yerokhin:2025:042820, Volkov:2025:preprint_03284}. In addition, we previously scaled the values of the one-electron two-loop contributions obtained for the Coulomb potential, when used them in calculations with screening potentials, see Ref.~\cite{Malyshev:2023:042806}. In this work, we have decided not to apply this scaling. 
We note that the resulting change in theoretical predictions is fully covered by our conservative estimate of the uncertainty associated with uncalculated higher-order contributions, which correspond to a set of one-electron two-loop diagrams with an additional photon line mediating the interelectronic interaction. This estimate is included in all uncertainties discussed below and is evaluated as the one-electron two-loop correction for the $1s$ state multiplied by a factor of $2/Z$~\cite{Malyshev:2024:062824}. 

The rigorous consideration in the first and second orders of the QED PT is further supplemented by the inclusion of several additional corrections. First, the electron-electron correlation effects of the third and higher orders are treated in the Breit approximation by means of the configuration-interaction (CI) method in the basis of the Dirac-Sturm orbitals~\cite{Bratzev:1977:2655, Tupitsyn:2003:022511}. 
It is well-known that the Dirac-Coulomb-Breit Hamiltonian~\cite{Faustov:1970:478, Sucher:1980:348, Mittleman:1981:1167}, which underlies the CI calculations, contains the projectors onto states constructed from positive-energy eigenfunctions of some chosen one-electron operator. Consequently, the Breit-approximation results, taken alone, generally vary depending on the specific decomposition into the positive- and negative-energy spectra~\cite{Sapirstein:1999:259}. This ambiguity can only be resolved within the framework of rigorous QED theory~\cite{Kozhedub:2019:062506, Malyshev:2023:042806}. In the present work, this decomposition is defined with respect to the same potential, which provides the initial approximation for the QED PT. This choice naturally follows from QED~\cite{Shabaev:1993:4703, Shabaev:2024:94:inbook} and ensures the most accurate merging with the QED calculations. The remaining minor ambiguity, manifesting itself in the higher-order electron-correlation contributions, is studied by comparing the results obtained for the LDF, CH, and pure Coulomb potentials, and is incorporated into the uncertainties.
Second, the higher-order screened QED contributions are estimated using the model-QED operator~\cite{Shabaev:2013:012513, Shabaev:2015:175:2018:69:join_pr}. 
Third, the contributions of the nuclear-recoil effect are taken into account. Namely, within the independent-electron approximation, the corresponding corrections are treated rigorously within QED to zeroth order in $\alpha$ and to all orders in $\alpha Z$ \cite{Shabaev:1985:394, Shabaev:1988:107, Shabaev:1998:59}; see also Refs.~\cite{Pachucki:1995:1854, Yelkhovsky:Budker, Adkins:2007:042508, Anisimova:2022:062823} and recent advances in Refs.~\cite{Pachucki:2024:032804, Pachucki:2025:032820}. The first-order interelectronic-interaction correction to the nuclear-recoil effect is evaluated within the Breit approximation based on the mass-shift Hamiltonian~\cite{Shabaev:1985:394, Shabaev:1988:107, Palmer:1987:5987}. 
Fourth, the corrections arising from the nuclear-polarization effect are taken into account as well \cite{Plunien:1995:1119:1996:4614:join_pr, Nefiodov:1996:227, Yerokhin:2015:033103}.

As in Ref.~\cite{Malyshev:2024:062824}, we consider nine Be-like ions, namely: argon~$^{40}{\rm Ar}^{14+}$, krypton~$^{84}{\rm Kr}^{32+}$, molybdenum~$^{98}{\rm Mo}^{38+}$, xenon~$^{132}{\rm Xe}^{50+}$, gold~$^{197}{\rm Au}^{75+}$, lead~$^{208}{\rm Pb}^{78+}$, bismuth~$^{209}{\rm Bi}^{79+}$, thorium~$^{232}{\rm Th}^{86+}$, and uranium~$^{238}{\rm U}^{88+}$.  The hyperfine
structure of gold and bismuth ions is neglected. The nuclear-charge distribution is described by the Fermi model with the thickness parameter equal to 2.3~fm. For $^{238}{\rm U}^{88+}$, we additionally take into account the nuclear-deformation effect following Ref.~\cite{Kozhedub:2008:032501}. The values of nuclear masses and root-mean-square radii are taken as in Ref.~\cite{Yerokhin:2015:033103}. The 2018 CODATA recommended values of the fundamental constants~\cite{Tiesinga:2021:025010} are used. 

\input{table_amplitude_J0.tex}

\input{table_amplitude_J1.tex}

\input{table_amplitude_J2.tex}

As noted above, the proximity of energy levels with the same symmetry in Be-like HCIs leads to their strong mixing due to the interelectronic interaction. In our approach, this issue is overcome by applying PT for quasidegenerate levels. The TTGF method implies combining a set of quasidegenerate levels into a finite-dimensional model subspace $\Omega$ and constructing an effective Hamiltonian~$H_{\rm eff}$, which acts on this subspace. The matrix of $H_{\rm eff}$ is evaluated step by step, incorporating all relevant contributions. When the matrix is obtained, its eigenvalues yield the desired energies of the mixed states. 

We denote the unperturbed many-electron wave functions by using the $jj$-coupling notations, while the states of interest in Be-like ions are referred to employing the $LS$-coupling scheme. This highlights, on the one hand, the fully relativistic nature of our approach and, on the other hand, the fact that the resulting states arise from level mixing. In the case of well-separated levels, where the model subspace $\Omega$ is one-dimensional and the matrix element of~$H_{\rm eff}$ directly yields the state energy, the correspondence between the two couplings is unambiguous. Such states include $2s2p \, ^3P_0$, $2s2p \, ^3P_2$, and $2p2p\,^3P_1$. In the $jj$ coupling, they correspond to the $(2s2p_{1/2})_0$, $(2s2p_{3/2})_2$, and $(2p_{1/2}2p_{3/2})_1$ states, respectively. All the other cases requires treatment using non-trivial model subspaces with dimensions greater than one. It is convenient to classify these states by the value of the total angular momentum~$J$. The $J=0$ states, $2s2s \, ^1S_0$, $2p2p\,^3P_0$, and $2p2p\,^1S_0$, are studied using the three-dimensional model subspace~$\Omega_{0}$ spanned by the $(2s2s)_0$, $(2p_{1/2}2p_{1/2})_0$, and $(2p_{3/2}2p_{3/2})_0$ levels. The $J=1$ states, $2s2p \, ^3P_1$ and $2s2p \, ^1P_1$, were considered in Ref.~\cite{Malyshev:2024:062824}. This consideration was based on the two-dimensional model subspace~$\Omega_{1}$ spanned by the $(2s2p_{1/2})_1$ and $(2s2p_{3/2})_1$ levels. Finally, the $J=2$ states, $2p2p\,^3P_2$ and $2p2p\,^1D_2$, are considered employing the two-dimensional model subspace~$\Omega_{2}$ spanned by the $(2p_{1/2}2p_{3/2})_2$ and $(2p_{3/2}2p_{3/2})_2$ levels.

To illustrate the extent to which the levels are mixed by the interelectronic interaction, in Tables~\ref{tab:amplitude_J0}-\ref{tab:amplitude_J2} we present examples of the CI calculations for the $J=0$, $1$, and $2$ states, respectively. Within the CI method, the Dirac-Coulomb-Breit equation is solved using the Ritz variational principle in a space of configuration-state functions (CSFs) with given values of the total angular momentum $J$ and its projection $M_J$. Tables~\ref{tab:amplitude_J0}-\ref{tab:amplitude_J2} show the expansion coefficients associated with the CSFs belonging to the model subspaces~$\Omega_J$. We note that the coefficients $A_{11}$, $A_{12}$, and $A_{13}$ from Table~\ref{tab:amplitude_J0} as well as those from Table~\ref{tab:amplitude_J1} were previously reported in Ref.~\cite{Malyshev:2024:062824}. They are tabulated here for completeness. 

The weight of a specific CSF can be obtained by taking the square of the corresponding expansion coefficient. In Tables~\ref{tab:amplitude_J0}-\ref{tab:amplitude_J2}, the sums of the coefficient squares are slightly less than one. For all ions, except for argon, the deviations do not exceed $5\times 10^{-4}$, while for $Z=18$ a maximum deviation of $2\times 10^{-3}$ occurs for the $2s2p \, ^1P_1$ state. These deviations stem from the contribution of CSFs lying beyond the model subspaces. If required, it is feasible, in principle, to enlarge the model subspaces for Be-like argon in order to enable even more precise studies of the electronic structure.

All the coefficients presented in Tables~\ref{tab:amplitude_J0}-\ref{tab:amplitude_J2} are obtained within a certain CI calculation for the LDF potential, which includes approximately $200\,000$ CSFs and prior to any extrapolation to the infinite-dimensional configuration space~\cite{Kaygorodov:2019:032505}. Consequently, they can not be regarded as the exact ones. Nevertheless, they provide a clear idea about the degree of level mixing. It can be seen that, for all the states, the coefficients corresponding to the dominant levels, $A_{ii}$, $B_{ii}$, and $C_{ii}$ (with $i=1,2,\ldots$), increase and approach one as the nuclear charge~$Z$ grows. This behavior reflects the fact that the quasidegeneracy of the levels is gradually lifted with increasing $Z$. Nevertheless, the extent of mixing depends on the states under consideration. For example, the mixing of the $J=0$ states remains substantial even for $Z=92$, whereas the mixing of the $J=2$ states becomes negligible in this case. As we have checked, application of the single-level QED approach for the $2p2p\,^3P_2$ and $2p2p\,^1D_2$ states in Be-like uranium yields the results which are consistent with those obtained within PT for quasidegenerate levels.


\section{Numerical results and discussions \label{sec:2}}

\input{table_excitation_sp.tex}

In this section, we present the results of our QED calculations of the excitation and transition energies in Be-like HCIs. All obtained theoretical predictions are accompanied by thoroughly analyzed uncertainties. The procedure used to estimate the uncertainties is described in details in Ref.~\cite{Malyshev:2024:062824}. The only change in this procedure is related to the updates in the evaluation of the one-electron two-loop contributions~\cite{Yerokhin:2025:042820, Volkov:2025:preprint_03284}. 

Before turning to the intra-$L$-shell doubly excited states, which are the primary focus of the present work, we first examine how the revisited treatment of the one-electron two-loop contributions, which has been discussed in the previous section, affects the energies of the previously studied singly excited states. The updated excitation energies of the $2s2p \, ^3P_{0,1,2}$ and $2s2p \, ^1P_1$ states from the $2s2s \, ^1S_0$ ground state are presented in Table~\ref{tab:excitation_sp}. By comparing with the corresponding values in Table~III in Ref.~\cite{Malyshev:2024:062824}, one can see that for low-$Z$ Be-like ions, where the total theoretical uncertainties are determined mainly by uncalculated higher-order QED effects, there are only minor changes in the last significant digits. In contrast, for high-$Z$ ions, the total theoretical uncertainties are reduced. In the following, these updated values are used when discussing transition energies. 

Our theoretical predictions for the excitation energies of the $2p2p\,^3P_{0,1,2}$, $2p2p\,^1D_2$, and $2p2p\,^1S_0$ states from the $2s2s\,^1S_0$ ground state, $2p2p\,^3P \rightarrow 2s2p\,^3P$ transition energies, and a number of other selected transition energies are compiled in Tables~\ref{tab:excitation_pp}, \ref{tab:2p2p_3P-2s2p_3P}, and \ref{tab:transition_pp}, respectively. These tables provide a detailed comparison of our results with the previous relativistic calculations and available experimental data. The previous theoretical studies of Be-like HCIs show a considerable scatter of the reported values. The lack of theoretical uncertainties in these works complicates comparisons, but all the results generally agree with each other. However, our predictions are much more precise, as confirmed by the comparison with the measurements. For instance, a perfect agreement is found between our value of $533.7306(68)$ for the $2p2p\,^3P_1 \rightarrow 2s2p\,^3P_0$ transition energy in Be-like xenon and the most recent, though less precise, experimental value of $533.733(22)$ obtained employing the resonant electron-ion collision process of dielectronic recombination~\cite{Bernhardt:2015:144008}.

In Table~\ref{tab:QED_pp}, we present a separation of the obtained theoretical predictions for the energies of the intra-$L$-shell doubly excited $2p2p\,^3P_{0,1,2}$, $2p2p\,^1D_2$, and $2p2p\,^1S_0$ states in Be-like HCIs into the non-QED and QED parts. Following the approach of Ref.~\cite{Malyshev:2024:062824}, the non-QED part is evaluated by diagonalizing the matrix of $H_{\rm eff}$, which incorporates only the results of the CI calculations, the correction due to the frequency dependence of the interelectronic-interaction operator, the non-QED part of the nuclear-recoil effect, and the nuclear-polarization correction. The QED part corresponds to the remainder and is calculated by subtracting the non-QED part from the total result. The separation shown in Table~\ref{tab:QED_pp} is based on the calculations for the LDF potential. Due to the mixing of levels, the individual contributions cease to be additive, making a further separation of the different terms impractical.
 

\section{Summary \label{sec:3}}

The present work completes the \textit{ab initio} QED treatment of the intra-$L$-shell excitations in Be-like ions by examining the doubly excited $2p2p\,^3P_{0,1,2}$, $2p2p\,^1D_2$, and $2p2p\,^1S_0$ states. The singly excited $2s2p \, ^3P_{0,1,2}$ and $2s2p \, ^1P_1$ states were investigated in our previous work~\cite{Malyshev:2024:062824}. The present calculations are performed for selected ions in a wide range: from ${\rm Ar}^{14+}$ to ${\rm U}^{88+}$. All the excitation energies are obtained relative to the $2s2s \, ^1S_0$ ground state. To properly take into account the mixing of close levels with the same symmetry, the QED perturbative approach for quasi-degenerate levels is used. Namely, this method is employed for the even states with the total angular momentum equal to zero, $2s2s \, ^1S_0$, $2p2p\,^3P_0$, and $2p2p\,^1S_0$, and for the even states with the total angular momentum equal to two, $2p2p\,^3P_2$ and $2p2p\,^1D_2$. Previously, the same approach was applied to the QED calculations of the odd states having the total angular momentum equal to one, $2s2p \, ^3P_1$ and $2s2p \, ^1P_1$. The remaining states, $2p2p\,^3P_1$ here and $2s2p \, ^3P_0$ and $2s2p \, ^3P_2$ previously, are treated using the standard QED perturbation theory for a single level. Our approach combines the first- and second-order QED contributions evaluated within the Furry picture and the third- and higher-order interelectronic-interaction corrections calculated within the Breit approximation. The model-QED operator is employed to estimate the higher-order screened QED effects. The nuclear-recoil and nuclear-polarization effects are taken into account as well. The detailed analysis of uncertainties associated with uncalculated effects is carried out, ensuring reliable error estimates. As a result, the most accurate theoretical predictions to date for the intra-$L$-shell excitation and transition energies in Be-like ions are obtained, which are in agreement with the available experimental data. The present calculations, in conjunction with those performed in Ref.~\cite{Malyshev:2024:062824}, provide a benchmark for future high-precision experimental and theoretical investigations of Be-like highly charged ions and establish a solid foundation for further rigorous tests of the bound-state-QED methods.


\section*{Acknowledgments}

The work was supported by the Russian Science Foundation (Grant No. 22-62-00004, https://rscf.ru/project/22-62-00004/).
We thank Vladimir Yerokhin for sharing with us the recent results of the calculations of one-electron two-loop contributions.
Calculations of higher-order electron-electron correlation effects were performed on the basis of the HybriLIT heterogeneous computing platform (LIT, JINR).


\newpage


\input{table_excitation_pp.tex}

\newpage
\input{table_2p2p_3P-2s2p_3P.tex}

\newpage
\input{table_transition_pp.tex}

\newpage
\input{table_QED_pp.tex}


\end{document}

%% file: table_amplitude_J0.tex
\begin{table*}[t]
\centering

\renewcommand{\arraystretch}{1.25}

\caption{\label{tab:amplitude_J0} 
         Expansion coefficients, $A_{ik}$, of the many-electron wave functions in the configuration-state functions:
         $\Psi[2s2s\,^1S_0]=A_{11}\Phi[(2s2s)_0]+A_{12}\Phi[(2p_{1/2}2p_{1/2})_0]+A_{13}\Phi[(2p_{3/2}2p_{3/2})_0]+\ldots\,$,
         $\Psi[2p2p\,^3P_0]=A_{21}\Phi[(2s2s)_0]+A_{22}\Phi[(2p_{1/2}2p_{1/2})_0]+A_{23}\Phi[(2p_{3/2}2p_{3/2})_0]+\ldots\,$, and
         $\Psi[2p2p\,^1S_0]=A_{31}\Phi[(2s2s)_0]+A_{32}\Phi[(2p_{1/2}2p_{1/2})_0]+A_{33}\Phi[(2p_{3/2}2p_{3/2})_0]+\ldots\,$.
         The configuration weights are equal to the squares of the coefficients.
         The coefficients are obtained by means of the configuration-interaction method for a given configuration space. 
         The decomposition into the positive- and negative-energy spectra is determined by the Dirac Hamiltonian with the local Dirac-Fock potential included.
         }
         
\begin{tabular}{
                l@{\qquad}
                S[table-format=2.4]                
                S[table-format=2.4] 
                S[table-format=2.4]@{\qquad} 
                S[table-format=-2.4]                
                S[table-format=2.4] 
                S[table-format=-2.4]@{\qquad} 
                S[table-format=-2.4]                
                S[table-format=2.4] 
                S[table-format=2.4]
               }
               
\hline
\hline

   \multirow{2}{*}{$Z$}    &
   \multicolumn{3}{c}{\rule{0pt}{1.2em} $2s2s\,^1S_0$~~~~~~}                             &
   \multicolumn{3}{c}{\rule{0pt}{1.2em} $2p2p\,^3P_0$~~~~~~}                              &
   \multicolumn{3}{c}{\rule{0pt}{1.2em} $2p2p\,^1S_0$}                              \\
                                                         &
   \multicolumn{1}{c}{\rule[-0.4em]{0pt}{0.4em} $A_{11}$}                                                         &
   \multicolumn{1}{c}{\rule[-0.4em]{0pt}{0.4em} $A_{12}$}                                                          &
   \multicolumn{1}{c}{\rule[-0.4em]{0pt}{0.4em} $A_{13}$~~~~~~}                                                          &
   \multicolumn{1}{c}{\rule[-0.4em]{0pt}{0.4em} $A_{21}$}                                                         &
   \multicolumn{1}{c}{\rule[-0.4em]{0pt}{0.4em} $A_{22}$}                                                          &
   \multicolumn{1}{c}{\rule[-0.4em]{0pt}{0.4em} $A_{23}$~~~~~~}                                                          &
   \multicolumn{1}{c}{\rule[-0.4em]{0pt}{0.4em} $A_{31}$}                                                         &
   \multicolumn{1}{c}{\rule[-0.4em]{0pt}{0.4em} $A_{32}$}                                                          &
   \multicolumn{1}{c}{\rule[-0.4em]{0pt}{0.4em} $A_{33}$}                                                          \\
        
\hline   
                       
  18 \rule{0pt}{3.2ex}  &     0.9739 &     0.1371 &     0.1800 &    -0.0292 &     0.8648 &    -0.5008 &    -0.2241 &     0.4821 &     0.8459    \\ 
  36  &     0.9840 &     0.1378 &     0.1126 &    -0.1153 &     0.9757 &    -0.1859 &    -0.1354 &     0.1698 &     0.9759    \\ 
  42  &     0.9864 &     0.1375 &     0.0897 &    -0.1261 &     0.9843 &    -0.1227 &    -0.1051 &     0.1096 &     0.9883    \\ 
  54  &     0.9895 &     0.1337 &     0.0550 &    -0.1306 &     0.9897 &    -0.0577 &    -0.0621 &     0.0499 &     0.9967    \\ 
  79  &     0.9930 &     0.1162 &     0.0209 &    -0.1159 &     0.9931 &    -0.0164 &    -0.0227 &     0.0138 &     0.9996    \\ 
  82  &     0.9933 &     0.1137 &     0.0188 &    -0.1135 &     0.9934 &    -0.0143 &    -0.0203 &     0.0120 &     0.9997    \\ 
  83  &     0.9934 &     0.1129 &     0.0182 &    -0.1127 &     0.9935 &    -0.0137 &    -0.0196 &     0.0115 &     0.9997    \\ 
  90  &     0.9940 &     0.1081 &     0.0143 &    -0.1080 &     0.9940 &    -0.0102 &    -0.0153 &     0.0085 &     0.9998    \\ 
  92  &     0.9942 &     0.1070 &     0.0133 &    -0.1069 &     0.9942 &    -0.0094 &    -0.0143 &     0.0078 &     0.9998    \\ 

\hline
\hline

\end{tabular}%

\end{table*}

%% file: table_amplitude_J1.tex
\begin{table}[t]
\centering

\renewcommand{\arraystretch}{1.25}

\caption{\label{tab:amplitude_J1} 
         Expansion coefficients, $B_{ik}$, of the many-electron wave functions in the configuration-state functions:
         $\Psi[2s2p\,^3P_1]=B_{11}\Phi[(2s2p_{1/2})_1]+B_{12}\Phi[(2s2p_{3/2})_1]+\ldots\,$ and
         $\Psi[2s2p\,^1P_1]=B_{21}\Phi[(2s2p_{1/2})_1]+B_{22}\Phi[(2s2p_{3/2})_1]+\ldots\,$.
         }
         
\begin{tabular}{
                l@{\qquad}
                S[table-format=2.4,group-separator=]                
                S[table-format=-2.4,group-separator=]@{\qquad}               
                S[table-format=2.4,group-separator=] 
                S[table-format=2.4,group-separator=]
               }
               
\hline
\hline

   \multirow{2}{*}{$Z$}    &
   \multicolumn{2}{c}{\rule{0pt}{1.2em} $2s2p\,^3P_1$~~~~~~}                             &
   \multicolumn{2}{c}{\rule{0pt}{1.2em} $2s2p\,^1P_1$}                              \\
                                                         &
   \multicolumn{1}{c}{\rule[-0.4em]{0pt}{0.4em} $B_{11}$}                                                         &
   \multicolumn{1}{c}{\rule[-0.4em]{0pt}{0.4em} $B_{12}$~~~~~~}                                                          &
   \multicolumn{1}{c}{\rule[-0.4em]{0pt}{0.4em} $B_{21}$}                                                         &
   \multicolumn{1}{c}{\rule[-0.4em]{0pt}{0.4em} $B_{22}$}                                                          \\
        
\hline   
                       
  18 \rule{0pt}{3.2ex}  &     0.8461 &    -0.5326 &     0.5321 &     0.8454    \\ 
  36  &     0.9602 &    -0.2790 &     0.2789 &     0.9601    \\ 
  42  &     0.9799 &    -0.1992 &     0.1991 &     0.9798    \\ 
  54  &     0.9948 &    -0.1016 &     0.1015 &     0.9947    \\ 
  79  &     0.9995 &    -0.0312 &     0.0312 &     0.9995    \\ 
  82  &     0.9996 &    -0.0275 &     0.0275 &     0.9996    \\ 
  83  &     0.9996 &    -0.0264 &     0.0264 &     0.9996    \\ 
  90  &     0.9998 &    -0.0200 &     0.0200 &     0.9998    \\ 
  92  &     0.9998 &    -0.0186 &     0.0185 &     0.9998    \\ 

\hline
\hline

\end{tabular}%

\end{table}

%% file: table_amplitude_J2.tex
\begin{table}[t]
\centering

\renewcommand{\arraystretch}{1.25}

\caption{\label{tab:amplitude_J2} 
         Expansion coefficients, $C_{ik}$, of the many-electron wave functions in the configuration-state functions:
         $\Psi[2p2p\,^3P_2]=C_{11}\Phi[(2p_{1/2}2p_{3/2})_2]+C_{12}\Phi[(2p_{3/2}2p_{3/2})_2]+\ldots\,$ and
         $\Psi[2p2p\,^1D_2]=C_{21}\Phi[(2p_{1/2}2p_{3/2})_2]+C_{22}\Phi[(2p_{3/2}2p_{3/2})_2]+\ldots\,$.
         }
         
\begin{tabular}{
                l@{\qquad}
                S[table-format=2.5,group-separator=]                
                S[table-format=2.5,group-separator=]@{\qquad}               
                S[table-format=-2.5,group-separator=] 
                S[table-format=2.5,group-separator=]
               }
               
\hline
\hline

   \multirow{2}{*}{$Z$}    &
   \multicolumn{2}{c}{\rule{0pt}{1.2em} $2p2p\,^3P_2$~~~~~~}                             &
   \multicolumn{2}{c}{\rule{0pt}{1.2em} $2p2p\,^1D_2$}                              \\
                                                         &
   \multicolumn{1}{c}{\rule[-0.4em]{0pt}{0.4em} $C_{11}$}                                                         &
   \multicolumn{1}{c}{\rule[-0.4em]{0pt}{0.4em} $C_{12}$~~~~~~}                                                          &
   \multicolumn{1}{c}{\rule[-0.4em]{0pt}{0.4em} $C_{21}$}                                                         &
   \multicolumn{1}{c}{\rule[-0.4em]{0pt}{0.4em} $C_{22}$}                                                          \\
        
\hline   
                       
  18 \rule{0pt}{3.2ex}  &    0.72217 &    0.69135 &   -0.69070 &    0.72188    \\ 
  36  &    0.98875 &    0.14862 &   -0.14846 &    0.98878    \\ 
  42  &    0.99587 &    0.08957 &   -0.08945 &    0.99590    \\ 
  54  &    0.99915 &    0.03931 &   -0.03924 &    0.99917    \\ 
  79  &    0.99990 &    0.01112 &   -0.01109 &    0.99991    \\ 
  82  &    0.99991 &    0.00979 &   -0.00975 &    0.99992    \\ 
  83  &    0.99992 &    0.00939 &   -0.00935 &    0.99993    \\ 
  90  &    0.99994 &    0.00706 &   -0.00703 &    0.99995    \\ 
  92  &    0.99994 &    0.00652 &   -0.00649 &    0.99995    \\ 

\hline
\hline

\end{tabular}%

\end{table}

%% file: table_excitation_sp.tex
\begin{table*}[t]
\centering

\renewcommand{\arraystretch}{1.25}

\caption{\label{tab:excitation_sp} 
         The excitation energies of the $2s2p\,^3P_{0,1,2}$ and $2s2p\,^1P_1$ states from the $2s2s\,^1S_0$ ground state in Be-like ions (in eV).
         Compared to Ref.~\cite{Malyshev:2024:062824}, the recent progress~\cite{Yerokhin:2025:042820,Volkov:2025:preprint_03284} in the calculations of the one-electron two-loop contributions is taken into account. 
         In addition, the scaling procedure previously applied to these corrections is no longer used.
         See the text for the details.
         }
         
\begin{tabular}{
                  l@{\qquad}
                  S[table-format=3.5(2),group-separator=]                
                  S[table-format=3.5(2),group-separator=] 
                  S[table-format=3.5(2),group-separator=] 
                  S[table-format=3.5(2),group-separator=]
               }
               
\hline
\hline
   
   \multirow{2}{*}{$Z$}    &
   \multicolumn{1}{c}{\parbox{1.7cm}{\centering \rule{0pt}{1.2em} $2s2p\,^3P_0$}~~~~~~}                             &
   \multicolumn{1}{c}{\parbox{1.7cm}{\centering \rule{0pt}{1.2em} $2s2p\,^3P_1$}~~~~~~}                              &
   \multicolumn{1}{c}{\parbox{1.7cm}{\centering \rule{0pt}{1.2em} $2s2p\,^3P_2$}~~~~~~}                              &
   \multicolumn{1}{c}{\parbox{1.7cm}{\centering \rule{0pt}{1.2em} $2s2p\,^1P_1$}~~~~~~}                              \\                                                                                
   
      &
   \multicolumn{1}{c}{\parbox{1.7cm}{\centering \rule[-0.4em]{0pt}{0.4em} $- 2s2s\,^1S_0$}~~~~~~}                   &
   \multicolumn{1}{c}{\parbox{1.7cm}{\centering \rule[-0.4em]{0pt}{0.4em} $- 2s2s\,^1S_0$}~~~~~~}                    &
   \multicolumn{1}{c}{\parbox{1.7cm}{\centering \rule[-0.4em]{0pt}{0.4em} $- 2s2s\,^1S_0$}~~~~~~}                    &
   \multicolumn{1}{c}{\parbox{1.7cm}{\centering \rule[-0.4em]{0pt}{0.4em} $- 2s2s\,^1S_0$}~~~~~~}                   \\
                                                                                                                 
\hline
     
  18 \rule{0pt}{3.2ex}  &       28.35405(41)  &        29.24429(59)  &        31.32955(39)  &        56.06801(69)    \\ 
  36  &        62.6308(21)  &         72.9864(20)  &        125.6573(18)  &        170.4201(21)    \\ 
  42  &        75.2880(32)  &         90.0053(32)  &        197.9865(29)  &        248.4990(30)    \\ 
  54  &       104.5314(72)  &        127.3007(72)  &        469.4829(68)  &        532.8006(67)    \\ 
  79  &        193.944(40)  &         229.650(39)  &        2191.808(38)  &        2289.598(37)    \\ 
  82  &        208.077(47)  &         244.946(47)  &        2584.793(45)  &        2687.575(45)    \\ 
  83  &        212.960(53)  &         250.192(53)  &        2728.851(51)  &        2833.347(51)    \\ 
  90  &         248.00(13)  &          287.38(13)  &         3951.40(13)  &         4068.63(13)    \\ 
  92  &        258.074(91)  &         297.915(90)  &        4380.634(87)  &        4501.772(86)    \\ 

\hline
\hline

\end{tabular}%

\end{table*}

%% file: table_excitation_pp.tex
\begin{table*}[t]
\centering

\renewcommand{\arraystretch}{1.25}

\caption{\label{tab:excitation_pp} 
         The excitation energies of the $2p2p\,^3P_{0,1,2}$, $2p2p\,^1D_2$, and $2p2p\,^1S_0$ states from the $2s2s\,^1S_0$ ground state in Be-like ions (in eV).
         The theoretical (Th.) results are compared with the experimental (Expt.) values.
         }
         
\begin{tabular}{
                  S[table-format=3.5(2),group-separator=]                
                  S[table-format=3.5(2),group-separator=] 
                  S[table-format=3.5(2),group-separator=] 
                  S[table-format=3.5(2),group-separator=]
                  S[table-format=3.5(2),group-separator=]
                  l@{\quad}
                  l@{\quad}
                  l
               }
               
\hline
\hline

   \multicolumn{1}{c}{\parbox{1.8cm}{\centering \rule{0pt}{1.2em} $2p2p\,^3P_0$}~~~~~~}                             &
   \multicolumn{1}{c}{\parbox{1.8cm}{\centering \rule{0pt}{1.2em} $2p2p\,^3P_1$}~~~~~~}                              &
   \multicolumn{1}{c}{\parbox{1.8cm}{\centering \rule{0pt}{1.2em} $2p2p\,^3P_2$}~~~~~~}                              &
   \multicolumn{1}{c}{\parbox{1.8cm}{\centering \rule{0pt}{1.2em} $2p2p\,^1D_2$}~~~~~~}                              &
   \multicolumn{1}{c}{\parbox{1.8cm}{\centering \rule{0pt}{1.2em} $2p2p\,^1S_0$}~~~~~~}                              &
   Th./                             &
   \multirow{2}{*}{Year}                                                                                         &
   \multirow{2}{*}{Reference}                                                                                    \\                                                                                

   \multicolumn{1}{c}{\parbox{1.8cm}{\centering \rule[-0.4em]{0pt}{0.4em} $- 2s2s\,^1S_0$}~~~~~~}                   &
   \multicolumn{1}{c}{\parbox{1.8cm}{\centering \rule[-0.4em]{0pt}{0.4em} $- 2s2s\,^1S_0$}~~~~~~}                    &
   \multicolumn{1}{c}{\parbox{1.8cm}{\centering \rule[-0.4em]{0pt}{0.4em} $- 2s2s\,^1S_0$}~~~~~~}                    &
   \multicolumn{1}{c}{\parbox{1.8cm}{\centering \rule[-0.4em]{0pt}{0.4em} $- 2s2s\,^1S_0$}~~~~~~}                    &
   \multicolumn{1}{c}{\parbox{1.8cm}{\centering \rule[-0.4em]{0pt}{0.4em} $- 2s2s\,^1S_0$}~~~~~~}                    &
   Expt.                   &
                                                                                                                 &
                                                                                                                 \\
                                                                                                                 
\hline
     
  \multicolumn{8}{c}{\rule{0pt}{3.1ex} $Z=18$}   \\ 
        75.00577(63)  &        76.26579(77)  &        77.89962(63)  &        85.49796(96)  &        104.2195(10)  &   Th.     &  2025  &  This work    \\ 

    75.0125    &     76.2740     &    77.9070      &    85.4889      &    104.196      &   Th.     &  2015  &  Wang  \textit{et al.} \cite{Wang:2015:16}                     \\

    75.0227    &     76.2841     &    77.9226      &    85.4532      &    104.1800     &   Th.     &  2005  &  Gu \cite{Gu:2005:267}                                         \\       
    
    74.9968    &     76.2585     &    77.8921      &    85.4298      &    104.1444     &   Th.     &  1996  &  Safronova \textit{et al.} \cite{Safronova:1996:4036}          \\
    
    75.0056    &     76.2662     &    77.8983      &                 &                 & Th.$^\dagger$ &  1985  &  Edl\'en \cite{Edlen:1985:86}                              \\ 
    
    75.4636    &     76.7052     &    78.3633      &    86.9200      &    105.9454     &   Th.     &  1979  &  Cheng  \textit{et al.} \cite{Cheng:1979:111}                  \\
    
  75.0001(38)  &  76.2676(31)    &  77.9003(36)    &                 &  104.2236(89)   & Expt.$^\ddagger$ &  2010  &  Saloman \cite{Saloman:2010:033101}                     \\ 
    
  \multicolumn{8}{c}{\rule{0pt}{3.1ex} $Z=36$}   \\ 
        178.3809(27)  &        226.6016(22)  &        236.6614(26)  &        296.6476(22)  &        331.2483(25)  &   Th.     &  2025  &  This work    \\ 
    
    178.5149   &     226.7879    &    236.8460     &    296.8690     &    331.4639     &   Th.     &  2005  &  Gu \cite{Gu:2005:267}                                         \\      
    
    178.3629   &     226.5824    &    236.6276     &    296.6109     &    331.2061     &   Th.     &  1996  &  Safronova \textit{et al.} \cite{Safronova:1996:4036}          \\
    
    178.0909   &     226.7372    &                 &                 &                 & Th.$^\dagger$ &  1985  &  Edl\'en \cite{Edlen:1985:86}                              \\ 
    
    178.8462   &     226.7212    &    237.3029     &    297.1761     &    332.3467     &   Th.     &  1979  &  Cheng  \textit{et al.} \cite{Cheng:1979:111}                  \\
     
    178.30(12) &    226.54(12)   &   236.79(12)    &   296.48(12)    &  331.22(12)     & Expt.$^\ddagger$ &  1991  & Sugar and Musgrove \cite{Sugar:1991:859}                \\
    
  \multicolumn{8}{c}{\rule{0pt}{3.1ex} $Z=42$}   \\ 
        216.7829(41)  &        319.8357(33)  &        332.0127(36)  &        450.0987(31)  &        488.9127(33)  &   Th.     &  2025  &  This work    \\ 
    
    216.9730   &     320.1380    &    332.3130     &    450.4999     &    489.3100     &   Th.     &  2005  &  Gu \cite{Gu:2005:267}                                         \\     
    
    216.7575   &     319.8009    &    331.9648     &    450.0453     &    488.8597     &   Th.     &  1996  &  Safronova \textit{et al.} \cite{Safronova:1996:4036}          \\
    
    217.1689   &     319.7347    &    332.4971     &    450.2849     &    489.6602     &   Th.     &  1979  &  Cheng  \textit{et al.} \cite{Cheng:1979:111}                  \\
    
  \multicolumn{8}{c}{\rule{0pt}{3.1ex} $Z=54$}   \\ 
        301.8858(96)  &        638.2621(81)  &        653.6757(82)  &       1012.1531(80)  &       1060.3924(79)  &   Th.     &  2025  &  This work    \\ 
    
    302.168    &     638.941     &    654.362      &    1013.220     &    1061.460     &   Th.     &  2005  &  Gu \cite{Gu:2005:267}                                         \\ 
    
    301.829    &     638.157     &    653.561      &    1012.004     &    1060.258     &   Th.     &  1996  &  Safronova \textit{et al.} \cite{Safronova:1996:4036}          \\
    
    302.030    &     637.620     &    653.671      &    1011.554     &    1060.343     &   Th.     &  1979  &  Cheng  \textit{et al.} \cite{Cheng:1979:111}                  \\
  \multicolumn{8}{c}{\rule{0pt}{3.1ex} $Z=79$}   \\ 
         541.251(69)  &        2491.730(64)  &        2509.498(64)  &        4503.726(64)  &        4577.830(64)  &   Th.     &  2025  &  This work    \\ 
    
    541.380    &    2491.555     &   2509.313      &    4503.290     &    4577.468     &   Th.     &  1996  &  Safronova \textit{et al.} \cite{Safronova:1996:4036}          \\
    
    540.625    &    2489.417     &   2507.833      &    4500.679     &    4575.419     &   Th.     &  1979  &  Cheng  \textit{et al.} \cite{Cheng:1979:111}                  \\ 
  \multicolumn{8}{c}{\rule{0pt}{3.1ex} $Z=82$}   \\ 
         577.675(84)  &        2904.313(77)  &        2921.884(77)  &        5296.383(78)  &        5374.229(78)  &   Th.     &  2025  &  This work    \\ 

    576.735    &    2901.545     &   2919.759      &    5292.765     &    5371.266     &   Th.     &  1979  &  Cheng  \textit{et al.} \cite{Cheng:1979:111}                  \\
  \multicolumn{8}{c}{\rule{0pt}{3.1ex} $Z=83$}   \\ 
         590.222(96)  &        3055.094(90)  &        3072.573(90)  &        5586.804(90)  &        5665.930(90)  &   Th.     &  2025  &  This work    \\ 
    
    590.411    &    3054.907     &   3072.373      &    5586.298     &    5665.512     &   Th.     &  1996  &  Safronova \textit{et al.} \cite{Safronova:1996:4036}          \\
    
  \multicolumn{8}{c}{\rule{0pt}{3.1ex} $Z=90$}   \\ 
          680.82(25)  &         4325.74(24)  &         4342.17(24)  &         8049.23(24)  &         8137.80(24)  &   Th.     &  2025  &  This work    \\ 
    
    681.223    &    4325.564     &   4341.978      &    8048.575     &    8137.265     &   Th.     &  1996  &  Safronova \textit{et al.} \cite{Safronova:1996:4036}          \\
    
  \multicolumn{8}{c}{\rule{0pt}{3.1ex} $Z=92$}   \\ 
          707.20(17)  &         4768.79(16)  &         4784.79(16)  &         8913.11(16)  &         9004.54(16)  &   Th.     &  2025  &  This work    \\ 
    
    707.899    &    4768.873     &   4784.852      &    8912.674     &    9004.232     &   Th.     &  1996  &  Safronova \textit{et al.} \cite{Safronova:1996:4036}          \\
    
    706.436    &    4766.005     &   4782.625      &    8909.350     &    9001.509     &   Th.     &  1979  &  Cheng  \textit{et al.} \cite{Cheng:1979:111}                  \\

\hline
\hline

\end{tabular}%

\vspace{2mm}

\begin{flushleft}

$^\dagger$ Semiempirical prediction.

$^\ddagger$ Compilation of energy levels obtained by fitting to available lines.

\end{flushleft}

\end{table*}

%% file: table_2p2p_3P-2s2p_3P.tex
\begin{table*}[t]
\centering

\renewcommand{\arraystretch}{1.25}

\caption{\label{tab:2p2p_3P-2s2p_3P} 
         The $2p2p\,^3P \rightarrow 2s2p\,^3P$ transition energies in Be-like ions (in eV).
         The theoretical (Th.) results are compared with the experimental (Expt.) values.
         }
         
\begin{tabular}{@{}
                  S[table-format=4.5(2),group-separator=,table-align-text-post=false,table-space-text-post=]
                  S[table-format=4.5(2),group-separator=,table-align-text-post=false,table-space-text-post=]
                  S[table-format=4.5(2),group-separator=]
                  S[table-format=3.5(2),group-separator=]
                  S[table-format=3.5(2),group-separator=,table-align-text-post=false,table-space-text-post=]
                  S[table-format=3.5(2),group-separator=]@{\quad}
                  l@{\quad}
                  l@{\quad}
                  l@{}
               }
               
\hline
\hline

   \multicolumn{1}{c}{\parbox{1.6cm}{\centering \rule{0pt}{1.2em} $2p2p\,^3P_2$}~~~~~~}                             &
   \multicolumn{1}{c}{\parbox{1.6cm}{\centering \rule{0pt}{1.2em} $2p2p\,^3P_1$}~~~~~~}                             &
   \multicolumn{1}{c}{\parbox{1.6cm}{\centering \rule{0pt}{1.2em} $2p2p\,^3P_1$}~~~~~~}                             &
   \multicolumn{1}{c}{\parbox{1.6cm}{\centering \rule{0pt}{1.2em} $2p2p\,^3P_2$}~~~~~~}                             &
   \multicolumn{1}{c}{\parbox{1.6cm}{\centering \rule{0pt}{1.2em} $2p2p\,^3P_0$}~~~~~~}                             &
   \multicolumn{1}{c}{\parbox{1.6cm}{\centering \rule{0pt}{1.2em} $2p2p\,^3P_1$}~~~~~~}                             &
   Th./                             &
   \multirow{2}{*}{Year}                                                                                         &
   \multirow{2}{*}{Reference}                                                                                    \\                                                                                

   \multicolumn{1}{c}{\parbox{1.6cm}{\centering \rule[-0.4em]{0pt}{0.4em} $- 2s2p\,^3P_1$}~~~~~~}                   &
   \multicolumn{1}{c}{\parbox{1.6cm}{\centering \rule[-0.4em]{0pt}{0.4em} $- 2s2p\,^3P_0$}~~~~~~}                   &
   \multicolumn{1}{c}{\parbox{1.6cm}{\centering \rule[-0.4em]{0pt}{0.4em} $- 2s2p\,^3P_1$}~~~~~~}                   &
   \multicolumn{1}{c}{\parbox{1.6cm}{\centering \rule[-0.4em]{0pt}{0.4em} $- 2s2p\,^3P_2$}~~~~~~}                   &
   \multicolumn{1}{c}{\parbox{1.6cm}{\centering \rule[-0.4em]{0pt}{0.4em} $- 2s2p\,^3P_1$}~~~~~~}                   &
   \multicolumn{1}{c}{\parbox{1.6cm}{\centering \rule[-0.4em]{0pt}{0.4em} $- 2s2p\,^3P_2$}~~~~~~}                   &
   Expt.                   &
                                                                                                                 &
                                                                                                                 \\

\hline
       
  \multicolumn{9}{c}{\rule{0pt}{3.1ex} $Z=18$}   \\ 
        48.65533(55)  &        47.91174(51)  &        47.02150(42)  &        46.57007(44)  &        45.76149(54)  &        44.93623(55)  &   Th.     &  2025  &  This work    \\ 

    48.6561     &    47.9136     &    47.0231     &    46.5687     &    45.7616     &   44.9357     &   Th.     &  2015  &  Wang  \textit{et al.} \cite{Wang:2015:16}              \\

    48.6641     &    47.9168     &    47.0256     &    46.5775     &    45.7642     &   44.9390     &   Th.     &  2005  &  Gu \cite{Gu:2005:267}                                  \\   
    
    48.6524     &    47.9096     &    47.0188     &    46.5652     &    45.7571     &   44.9316     &   Th.     &  1996  &  Safronova \textit{et al.} \cite{Safronova:1996:4036}   \\
     
    48.6550     &    47.9142     &    47.0228     &    46.5696     &    45.7623     &   44.9374     & Th.$^\dagger$ &  1985  &  Edl\'en \cite{Edlen:1985:86}                       \\ 
     
    48.9233     &    48.1568     &    47.2651     &    46.8498     &    46.0236     &   45.1916     &   Th.     &  1979  &  Cheng \textit{et al.} \cite{Cheng:1979:111}            \\
     
    48.6548(38) &    47.9144(37) &    47.0232(36) &    46.5740(35) &    45.7572(34) &   44.9405(33) &  Expt.    &  1987  &  Stewart \textit{et al.} \cite{Stewart:1987:126}        \\
    
    48.6556(95) &    47.9129(93) &    47.0207(89) &    46.5721(87) &    45.7709(84) &   44.9348(81) &  Expt.    &  1980  &  Fawcett \textit{et al.} \cite{Fawcett:1980:1349}       \\
  \multicolumn{9}{c}{\rule{0pt}{3.1ex} $Z=36$}   \\ 
        163.6750(21)  &        163.9708(18)  &        153.6153(19)  &        111.0041(24)  &        105.3945(21)  &        100.9444(20)  &   Th.     &  2025  &  This work    \\ 

   163.7813     &   164.0909     &   153.7232     &   111.0730     &   105.4502     &  101.0149     &   Th.     &  2005  &  Gu \cite{Gu:2005:267}                                  \\
   
   163.6524     &   163.9625     &   153.6072     &   110.9867     &   105.3878     &  100.9415     &   Th.     &  1996  &  Safronova \textit{et al.} \cite{Safronova:1996:4036}   \\
   
                &   164.22       &   153.85       &                &   105.21       &  101.1        & Th.$^\dagger$ &  1985  &  Edl\'en \cite{Edlen:1985:86}                       \\
                
   164.2840     &   164.2231     &   153.7023     &   111.8649     &   105.8273     &  101.2832     &   Th.     &  1979  &  Cheng \textit{et al.} \cite{Cheng:1979:111}            \\
   
   163.87(11)$^\ddagger$   &   163.87(11)$^\ddagger$   &   153.54(15)   &   111.05(5)    &   105.30(9)    &  100.72(16)   &  Expt.    &  1990  &  Martin \textit{et al.} \cite{Martin:1990:6570}         \\
  \multicolumn{9}{c}{\rule{0pt}{3.1ex} $Z=42$}   \\ 
        242.0074(30)  &        244.5478(29)  &        229.8304(29)  &        134.0263(35)  &        126.7776(33)  &        121.8493(32)  &   Th.     &  2025  &  This work    \\ 

   242.1961     &   244.7582     &   230.0211     &   134.1270     &   126.8561     &  121.9520     &   Th.     &  2005  &  Gu \cite{Gu:2005:267}                                  \\
   
   241.9754     &   244.5318     &   229.8114     &   134.0094     &   126.7680     &  121.8455     &   Th.     &  1996  &  Safronova \textit{et al.} \cite{Safronova:1996:4036}   \\ 
   
   242.5726     &   244.7789     &   229.8102     &   134.9868     &   127.2444     &  122.2244     &   Th.     &  1979  &  Cheng \textit{et al.} \cite{Cheng:1979:111}            \\
  \multicolumn{9}{c}{\rule{0pt}{3.1ex} $Z=54$}   \\ 
        526.3750(67)  &        533.7306(68)  &        510.9614(67)  &        184.1928(75)  &        174.5851(73)  &        168.7792(72)  &   Th.     &  2025  &  This work    \\ 

   526.887      &   534.278      &   511.466      &   184.358      &   174.693      &  168.937      &   Th.     &  2005  &  Gu \cite{Gu:2005:267}                                  \\
   
   526.294      &   533.675      &   510.890      &   184.175      &   174.562      &  168.771      &   Th.     &  1996  &  Safronova \textit{et al.} \cite{Safronova:1996:4036}   \\ 
   
   526.826      &   533.898      &   510.774      &   185.333      &   175.184      &  169.282      &   Th.     &  1979  &  Cheng \textit{et al.} \cite{Cheng:1979:111}            \\
   
                &   533.733(22)  &                &                &                &               &  Expt.    &  2015  &  Bernhardt \textit{et al.} \cite{Bernhardt:2015:144008} \\
                
                &                &                &                &   174.4$(1.2)$ &               &  Expt.    &  1988  &  Martin \textit{et al.} \cite{Martin:1988:79} \\ 
  \multicolumn{9}{c}{\rule{0pt}{3.1ex} $Z=79$}   \\ 
        2279.849(38)  &        2297.787(38)  &        2262.080(38)  &         317.690(39)  &         311.601(40)  &         299.922(39)  &   Th.     &  2025  &  This work    \\ 

  2279.631      &  2297.632      &  2261.873      &   317.836      &   311.698      &  300.078      &   Th.     &  1996  &  Safronova \textit{et al.} \cite{Safronova:1996:4036}   \\ 
   
  2280.198      &  2297.806      &  2261.783      &   319.453      &   312.991      &  301.037      &   Th.     &  1979  &  Cheng \textit{et al.} \cite{Cheng:1979:111}            \\
  \multicolumn{9}{c}{\rule{0pt}{3.1ex} $Z=82$}   \\ 
        2676.938(45)  &        2696.235(45)  &        2659.367(45)  &         337.091(47)  &         332.729(48)  &         319.520(47)  &   Th.     &  2025  &  This work    \\ 

  2677.220      &  2696.167      &  2659.006      &   338.881      &   334.196      &  320.667      &   Th.     &  1979  &  Cheng \textit{et al.} \cite{Cheng:1979:111}            \\
  \multicolumn{9}{c}{\rule{0pt}{3.1ex} $Z=83$}   \\ 
        2822.381(51)  &        2842.134(51)  &        2804.902(51)  &         343.722(53)  &         340.030(54)  &         326.243(53)  &   Th.     &  2025  &  This work    \\ 

  2822.131      &  2841.954      &  2804.665      &   343.914      &   340.169      &  326.448      &   Th.     &  1996  &  Safronova \textit{et al.} \cite{Safronova:1996:4036}   \\ 
  \multicolumn{9}{c}{\rule{0pt}{3.1ex} $Z=90$}   \\ 
         4054.79(13)  &         4077.73(13)  &         4038.35(13)  &          390.77(13)  &          393.43(13)  &          374.34(13)  &   Th.     &  2025  &  This work    \\ 

  4054.467      &  4077.492      &  4038.053      &   391.104      &   393.712      &  374.690      &   Th.     &  1996  &  Safronova \textit{et al.} \cite{Safronova:1996:4036}   \\ 
  \multicolumn{9}{c}{\rule{0pt}{3.1ex} $Z=92$}   \\ 
        4486.879(87)  &        4510.721(87)  &        4470.880(87)  &         404.160(90)  &         409.286(93)  &         388.161(90)  &   Th.     &  2025  &  This work    \\ 

  4486.675      &  4510.597      &  4470.696      &   404.654      &   409.722      &  388.675      &   Th.     &  1996  &  Safronova \textit{et al.} \cite{Safronova:1996:4036}   \\ 
   
  4488.033      &  4511.427      &  4471.413      &   406.840      &   411.844      &  390.219      &   Th.     &  1979  &  Cheng \textit{et al.} \cite{Cheng:1979:111}            \\

\hline
\hline

\end{tabular}%

\vspace{2mm}

\begin{flushleft}

$^\dagger$ Semiempirical prediction.

$^\ddagger$ Intensity is shared by these two lines.

\end{flushleft}

\end{table*}

%% file: table_transition_pp.tex
\begin{table*}[t]
\centering

\renewcommand{\arraystretch}{1.25}

\caption{\label{tab:transition_pp} 
         The selected transition energies in Be-like ions (in eV).
         The theoretical (Th.) results are compared with the experimental (Expt.) values.
         }
         
\resizebox{1.07\textwidth}{!}{%
\begin{tabular}{@{}
                  S[table-format=2.5(2),group-separator=]@{}
                  S[table-format=4.5(2),group-separator=]@{}
                  S[table-format=3.5(2),group-separator=]@{}
                  S[table-format=4.5(2),group-separator=,table-align-text-post=false,table-space-text-post=]@{}
                  S[table-format=4.5(2),group-separator=]@{}
                  S[table-format=4.5(2),group-separator=]@{}
                  S[table-format=4.5(2),group-separator=]@{\,\,\,}
                  l@{\,\,}
                  l@{\,\,}
                  l@{}
               }
               
\hline
\hline

   \multicolumn{1}{c}{\parbox{1.6cm}{\centering \rule{0pt}{1.2em} $2p2p\,^3P_2$}~~~}                             &
   \multicolumn{1}{c}{\parbox{1.6cm}{\centering \rule{0pt}{1.2em} $2p2p\,^3P_1$}~~~}                             &
   \multicolumn{1}{c}{\parbox{1.65cm}{\centering \rule{0pt}{1.2em} $2p2p\,^1S_0$}~~~}                             &
   \multicolumn{1}{c}{\parbox{1.6cm}{\centering \rule{0pt}{1.2em} $2p2p\,^1S_0$}~~~}                             &
   \multicolumn{1}{c}{\parbox{1.6cm}{\centering \rule{0pt}{1.2em} $2p2p\,^1D_2$}~~~}                             &
   \multicolumn{1}{c}{\parbox{1.6cm}{\centering \rule{0pt}{1.2em} $2p2p\,^1D_2$}~~~}                             &
   \multicolumn{1}{c}{\parbox{1.6cm}{\centering \rule{0pt}{1.2em} $2p2p\,^1D_2$}~~~}                             &
   Th./~~                             &
   \multirow{2}{*}{Year~}                                                                                         &
   \multirow{2}{*}{Reference}                                                                                    \\                                                                                

   \multicolumn{1}{c}{\parbox{1.6cm}{\centering \rule[-0.4em]{0pt}{0.4em} $- 2p2p\,^3P_1$}~~~}                   &
   \multicolumn{1}{c}{\parbox{1.6cm}{\centering \rule[-0.4em]{0pt}{0.4em} $- 2p2p\,^3P_0$}~~~}                   &
   \multicolumn{1}{c}{\parbox{1.65cm}{\centering \rule[-0.4em]{0pt}{0.4em} $- 2p2p\,^1D_2$}~~~}                   &
   \multicolumn{1}{c}{\parbox{1.6cm}{\centering \rule[-0.4em]{0pt}{0.4em} $- 2s2p\,^1P_1$}~~~}                   &
   \multicolumn{1}{c}{\parbox{1.6cm}{\centering \rule[-0.4em]{0pt}{0.4em} $- 2s2p\,^3P_1$}~~~}                   &
   \multicolumn{1}{c}{\parbox{1.6cm}{\centering \rule[-0.4em]{0pt}{0.4em} $- 2s2p\,^3P_2$}~~~}                   &
   \multicolumn{1}{c}{\parbox{1.6cm}{\centering \rule[-0.4em]{0pt}{0.4em} $- 2s2p\,^1P_1$}~~~}                   &
   Expt.~~                   &
                                                                                                                 &
                                                                                                                 \\

\hline
       
  \multicolumn{9}{c}{\rule{0pt}{2.9ex} $Z=18$}   \\ 
         1.63383(41)  &         1.26001(43)  &        18.72157(51)  &        48.15152(69)  &        56.25367(81)  &        54.16840(73)  &        29.42995(52)  &   Th.     &  2025  &  This work    \\ 

    1.6330   &   1.2615   &   18.707    &   48.125    &   56.2380   &   54.1506   &   29.4185   &   Th.     &  2015  &  Wang  \textit{et al.} \cite{Wang:2015:16}              \\

    1.6385   &   1.2614   &   18.7268   &   48.1532   &   56.1947   &   54.1081   &   29.4264   &   Th.     &  2005  &  Gu \cite{Gu:2005:267}                                  \\   
    
    1.6336   &   1.2617   &   18.7145   &   48.1572   &   56.1901   &   54.1030   &   29.4427   &   Th.     &  1996  &  Safronova \textit{et al.} \cite{Safronova:1996:4036}   \\
    
    1.6321   &   1.2605   &             &             &             &             &             & Th.$^\dagger$ &  1985  &  Edl\'en \cite{Edlen:1985:86}                       \\ 
    
    1.6582   &   1.2416   &   19.0254   &   48.2015   &   57.4799   &   55.4064   &   29.1761   &   Th.     &  1979  &  Cheng \textit{et al.} \cite{Cheng:1979:111}            \\
    
             &            &             &   48.366    &             &             &   29.384    &   Th.     &  1979  &  Glass \cite{Glass:1979:689}                            \\
             
             &            &             &   48.1602(37)$^\ddagger$ &           &             &             &  Expt.    &  1987  &  Stewart \textit{et al.} \cite{Stewart:1987:126}        \\

  \multicolumn{9}{c}{\rule{0pt}{2.9ex} $Z=36$}   \\ 
         10.0597(19)  &         48.2207(20)  &         34.6007(19)  &        160.8282(20)  &        223.6612(21)  &        170.9903(20)  &        126.2275(19)  &   Th.     &  2025  &  This work    \\ 

   10.0581   &  48.2730   &   34.5949   &  160.9140   &  223.8043   &  171.0960   &  126.3191   &   Th.     &  2005  &  Gu \cite{Gu:2005:267}                                  \\ 

   10.0452   &  48.2194   &   34.5952   &  160.8070   &  223.6358   &  170.9700   &  126.2118   &   Th.     &  1996  &  Safronova \textit{et al.} \cite{Safronova:1996:4036}   \\  
   
             &  48.6463   &             &             &             &             &             & Th.$^\dagger$ &  1985  &  Edl\'en \cite{Edlen:1985:86}                       \\ 
             
   10.5817   &  47.8750   &   35.1706   &  160.7895   &  224.1572   &  171.7381   &  125.6189   &   Th.     &  1979  &  Cheng \textit{et al.} \cite{Cheng:1979:111}            \\
   
   10.33(21) &  48.24(24) &   34.54(24) &  160.81(10) &             &  170.64(12) &  126.27(13) &  Expt.    &  1990  &  Martin \textit{et al.} \cite{Martin:1990:6570}         \\ 
    
  \multicolumn{9}{c}{\rule{0pt}{2.9ex} $Z=42$}   \\ 
         12.1770(29)  &        103.0528(32)  &         38.8140(30)  &        240.4136(30)  &        360.0933(31)  &        252.1122(30)  &        201.5996(29)  &   Th.     &  2025  &  This work    \\ 

   12.1750   & 103.1650   &   38.8100   &  240.5770   &  360.3830   &  252.3139   &  201.7669   &   Th.     &  2005  &  Gu \cite{Gu:2005:267}                                  \\ 

   12.1640   & 103.0434   &   38.8144   &  240.3766   &  360.0558   &  252.0899   &  201.5622   &   Th.     &  1996  &  Safronova \textit{et al.} \cite{Safronova:1996:4036}   \\  
             
   12.7624   & 102.5658   &   39.3753   &  240.3726   &  360.3604   &  252.7746   &  200.9974   &   Th.     &  1979  &  Cheng \textit{et al.} \cite{Cheng:1979:111}            \\
  \multicolumn{9}{c}{\rule{0pt}{2.9ex} $Z=54$}   \\ 
         15.4136(63)  &        336.3763(73)  &         48.2393(63)  &        527.5918(67)  &        884.8524(76)  &        542.6702(67)  &        479.3525(68)  &   Th.     &  2025  &  This work    \\ 

   15.421    & 336.773    &   48.240    &  528.059    &  885.745    &  543.216    &  479.819    &   Th.     &  2005  &  Gu \cite{Gu:2005:267}                                  \\ 

   15.404    & 336.328    &   48.254    &  527.499    &  884.737    &  542.618    &  479.245    &   Th.     &  1996  &  Safronova \textit{et al.} \cite{Safronova:1996:4036}   \\  
             
   16.051    & 335.590    &   48.789    &  527.577    &  884.708    &  543.216    &  478.788    &   Th.     &  1979  &  Cheng \textit{et al.} \cite{Cheng:1979:111}            \\
  \multicolumn{9}{c}{\rule{0pt}{2.9ex} $Z=79$}   \\ 
          17.768(22)  &        1950.479(31)  &          74.104(22)  &        2288.232(37)  &        4274.076(43)  &        2311.918(37)  &        2214.128(38)  &   Th.     &  2025  &  This work    \\ 

   17.758    & 1950.175  &   74.178   & 2288.027   & 4273.608   & 2311.813   & 2213.849   &   Th.     &  1996  &  Safronova \textit{et al.} \cite{Safronova:1996:4036}   \\  
             
   18.415    & 1948.792  &   74.740   & 2288.390   & 4273.045   & 2312.299   & 2213.650   &   Th.     &  1979  &  Cheng \textit{et al.} \cite{Cheng:1979:111}            \\   
  \multicolumn{9}{c}{\rule{0pt}{2.9ex} $Z=82$}   \\ 
          17.571(26)  &        2326.637(36)  &          77.846(26)  &        2686.654(45)  &        5051.437(52)  &        2711.591(45)  &        2608.808(45)  &   Th.     &  2025  &  This work    \\ 

    18.214    & 2324.810  &   78.501   & 2686.782   & 5050.226   & 2711.887   & 2608.281   &   Th.     &  1979  &  Cheng \textit{et al.} \cite{Cheng:1979:111}            \\  
    
  \multicolumn{9}{c}{\rule{0pt}{2.9ex} $Z=83$}   \\ 
          17.478(27)  &        2464.872(38)  &          79.126(27)  &        2832.583(51)  &        5336.612(58)  &        2857.953(51)  &        2753.457(51)  &   Th.     &  2025  &  This work    \\ 

   17.466    & 2464.496  &   79.214   & 2832.358   & 5336.056   & 2857.839   & 2753.144   &   Th.     &  1996  &  Safronova \textit{et al.} \cite{Safronova:1996:4036}   \\  
  \multicolumn{9}{c}{\rule{0pt}{2.9ex} $Z=90$}   \\ 
          16.433(37)  &        3644.917(55)  &          88.569(37)  &         4069.17(13)  &         7761.85(14)  &         4097.83(13)  &         3980.60(13)  &   Th.     &  2025  &  This work    \\ 

   16.414    & 3644.341  &   88.690   & 4068.905   & 7761.064   & 4097.701   & 3980.215   &   Th.     &  1996  &  Safronova \textit{et al.} \cite{Safronova:1996:4036}   \\  
  \multicolumn{9}{c}{\rule{0pt}{2.9ex} $Z=92$}   \\ 
          15.999(40)  &        4061.594(60)  &          91.427(40)  &        4502.765(86)  &        8615.196(97)  &        4532.477(86)  &        4411.339(86)  &   Th.     &  2025  &  This work    \\ 

   15.979    & 4060.974  &   91.558   & 4502.630   & 8614.497   & 4532.476   & 4411.072   &   Th.     &  1996  &  Safronova \textit{et al.} \cite{Safronova:1996:4036}   \\  
             
   16.620    & 4059.569  &   92.160   & 4503.901   & 8614.757   & 4533.564   & 4411.741   &   Th.     &  1979  &  Cheng \textit{et al.} \cite{Cheng:1979:111}            \\

\hline
\hline

\end{tabular}%
}

\vspace{2mm}

\begin{flushleft}

$^\dagger$ Semiempirical prediction.

$^\ddagger$ The line is blended with another one that may affect the measured wavelength.

\end{flushleft}

\end{table*}

%% file: table_QED_pp.tex
\begin{table*}[t]
\centering

\renewcommand{\arraystretch}{1.25}

\caption{\label{tab:QED_pp} 
         Non-QED and QED contributions to the excitation energies of the $2p2p\,^3P_{0,1,2}$, $2p2p\,^1D_2$, and $2p2p\,^1S_0$ states from the $2s2s\,^1S_0$ ground state in Be-like ions.
         See the text for details.
         }
         
\begin{tabular}{@{}
                  l
                  S[table-format=-2.5(2), group-separator=, table-space-text-pre=$<\,$]                
                  S[table-format=4.5(2), group-separator=, table-space-text-pre=$<\,$] 
                  S[table-format=4.5(2), group-separator=, table-space-text-pre=$<\,$] 
                  S[table-format=4.5(2), group-separator=, table-space-text-pre=$<\,$] 
                  S[table-format=4.4(2), group-separator=, table-space-text-pre=$<\,$] 
               }
               
\hline
\hline

   \multirow{2}{*}{Contribution}    &
   \multicolumn{1}{c}{\parbox{1.8cm}{\centering \rule{0pt}{1.2em} $2p2p\,^3P_0$}}                             &
   \multicolumn{1}{c}{\parbox{1.8cm}{\centering \rule{0pt}{1.2em} $2p2p\,^3P_1$}}                              &
   \multicolumn{1}{c}{\parbox{1.8cm}{\centering \rule{0pt}{1.2em} $2p2p\,^3P_2$}}                              &
   \multicolumn{1}{c}{\parbox{1.8cm}{\centering \rule{0pt}{1.2em} $2p2p\,^1D_2$}}                              &
   \multicolumn{1}{c}{\parbox{1.8cm}{\centering \rule{0pt}{1.2em} $2p2p\,^1S_0$}}                              \\
                                                   &
   \multicolumn{1}{c}{\parbox{1.8cm}{\centering \rule[-0.4em]{0pt}{0.4em} $- 2s2s\,^1S_0$}}                                                         &
   \multicolumn{1}{c}{\parbox{1.8cm}{\centering \rule[-0.4em]{0pt}{0.4em} $- 2s2s\,^1S_0$}}                                                          &
   \multicolumn{1}{c}{\parbox{1.8cm}{\centering \rule[-0.4em]{0pt}{0.4em} $- 2s2s\,^1S_0$}}                                                          &
   \multicolumn{1}{c}{\parbox{1.8cm}{\centering \rule[-0.4em]{0pt}{0.4em} $- 2s2s\,^1S_0$}}                                                          &
   \multicolumn{1}{c}{\parbox{1.8cm}{\centering \rule[-0.4em]{0pt}{0.4em} $- 2s2s\,^1S_0$}}                                                          \\ 
        
\hline
       
  \multicolumn{6}{c}{\rule{0pt}{3.1ex} $Z=18$}   \\ 
 $E_{\text{non-QED}}$  &      75.24390  &      76.50060  &      78.13138  &      85.72937  &      104.4416    \\ 
 $E_{\rm QED}$         &      -0.23812  &      -0.23481  &      -0.23176  &      -0.23141  &       -0.2220    \\ 
 $E_{\rm total}$ \rule[-0.6em]{0pt}{0.6em}      &  75.00577(63)  &  76.26579(77)  &  77.89962(63)  &  85.49796(96)  &  104.2195(10)    \\ 
  \multicolumn{6}{c}{\rule{0pt}{3.1ex} $Z=36$}   \\ 
 $E_{\text{non-QED}}$  &      181.2792  &      229.4177  &      239.4775  &      299.3411  &      333.9050    \\ 
 $E_{\rm QED}$         &       -2.8983  &       -2.8160  &       -2.8161  &       -2.6935  &       -2.6567    \\ 
 $E_{\rm total}$ \rule[-0.6em]{0pt}{0.6em}      &  178.3809(27)  &  226.6016(22)  &  236.6614(26)  &  296.6476(22)  &  331.2483(25)    \\ 
  \multicolumn{6}{c}{\rule{0pt}{3.1ex} $Z=42$}   \\ 
 $E_{\text{non-QED}}$  &      221.7600  &      324.6601  &      336.8403  &      454.6924  &      493.4690    \\ 
 $E_{\rm QED}$         &       -4.9771  &       -4.8243  &       -4.8276  &       -4.5938  &       -4.5563    \\ 
 $E_{\rm total}$ \rule[-0.6em]{0pt}{0.6em}      &  216.7829(41)  &  319.8357(33)  &  332.0127(36)  &  450.0987(31)  &  488.9127(33)    \\ 
  \multicolumn{6}{c}{\rule{0pt}{3.1ex} $Z=54$}   \\ 
 $E_{\text{non-QED}}$  &      313.8658  &      649.8307  &      665.2555  &     1023.1126  &     1071.3233    \\ 
 $E_{\rm QED}$         &      -11.9801  &      -11.5687  &      -11.5799  &      -10.9595  &      -10.9309    \\ 
 $E_{\rm total}$ \rule[-0.6em]{0pt}{0.6em}      &  301.8858(96)  &  638.2621(81)  &  653.6757(82)  &  1012.1531(80)  &  1060.3924(79)    \\ 
  \multicolumn{6}{c}{\rule{0pt}{3.1ex} $Z=79$}   \\ 
 $E_{\text{non-QED}}$  &       587.413  &      2536.495  &      2554.305  &      4546.480  &      4620.573    \\ 
 $E_{\rm QED}$         &       -46.162  &       -44.765  &       -44.807  &       -42.754  &       -42.743    \\ 
 $E_{\rm total}$ \rule[-0.6em]{0pt}{0.6em}      &   541.251(69)  &  2491.730(64)  &  2509.498(64)  &  4503.726(64)  &  4577.830(64)    \\ 
  \multicolumn{6}{c}{\rule{0pt}{3.1ex} $Z=82$}   \\ 
 $E_{\text{non-QED}}$  &       630.511  &      2955.679  &      2973.296  &      5345.604  &      5423.440    \\ 
 $E_{\rm QED}$         &       -52.836  &       -51.366  &       -51.413  &       -49.221  &       -49.211    \\ 
 $E_{\rm total}$ \rule[-0.6em]{0pt}{0.6em}      &   577.675(84)  &  2904.313(77)  &  2921.884(77)  &  5296.383(78)  &  5374.229(78)    \\ 
  \multicolumn{6}{c}{\rule{0pt}{3.1ex} $Z=83$}   \\ 
 $E_{\text{non-QED}}$  &       645.438  &      3108.827  &      3126.354  &      5638.357  &      5717.473    \\ 
 $E_{\rm QED}$         &       -55.216  &       -53.732  &       -53.781  &       -51.552  &       -51.543    \\ 
 $E_{\rm total}$ \rule[-0.6em]{0pt}{0.6em}      &   590.222(96)  &  3055.094(90)  &  3072.573(90)  &  5586.804(90)  &  5665.930(90)    \\ 
  \multicolumn{6}{c}{\rule{0pt}{3.1ex} $Z=90$}   \\ 
 $E_{\text{non-QED}}$  &        755.02  &       4398.60  &       4415.10  &       8119.91  &       8208.47    \\ 
 $E_{\rm QED}$         &        -74.20  &        -72.87  &        -72.93  &        -70.68  &        -70.67    \\ 
 $E_{\rm total}$ \rule[-0.6em]{0pt}{0.6em}      &    680.82(25)  &   4325.74(24)  &   4342.17(24)  &   8049.23(24)  &   8137.80(24)    \\ 
  \multicolumn{6}{c}{\rule{0pt}{3.1ex} $Z=92$}   \\ 
 $E_{\text{non-QED}}$  &        787.62  &       4848.04  &       4864.11  &       8990.27  &       9081.69    \\ 
 $E_{\rm QED}$         &        -80.42  &        -79.25  &        -79.31  &        -77.16  &        -77.15    \\ 
 $E_{\rm total}$ \rule[-0.6em]{0pt}{0.6em}      &    707.20(17)  &   4768.79(16)  &   4784.79(16)  &   8913.11(16)  &   9004.54(16)    \\ 

\hline
\hline

\end{tabular}%

\end{table*}